\documentclass[journal]{IEEEtran}
\usepackage[utf8]{inputenc}

\ifCLASSINFOpdf
   \usepackage[pdftex]{graphicx}
   \graphicspath{{figures/}}
   \DeclareGraphicsExtensions{.pdf,.jpeg,.png}
\else
   \usepackage[dvips]{graphicx}
   \graphicspath{{figures/}}
   \DeclareGraphicsExtensions{.eps}
\fi
\usepackage[cmex10]{amsmath}
\usepackage{array}
\usepackage[caption=false,font=footnotesize]{subfig}

\begin{document}
%
\title{\emph{Ab initio} simulation of band-to-band tunneling FETs with
  single- and few-layer 2-D materials as channels}
%
%
%
\author{
    \IEEEauthorblockN{Áron~Szabó\IEEEauthorrefmark{1},
      Cedric~Klinkert\IEEEauthorrefmark{1}, Davide
      Campi\IEEEauthorrefmark{2},
      Christian~Stieger\IEEEauthorrefmark{1},
      Nicola~Marzari\IEEEauthorrefmark{2}, and
    Mathieu~Luisier\IEEEauthorrefmark{1}}
\\\
    \IEEEauthorblockA{\IEEEauthorrefmark{1}Integrated Systems Laboratory, ETH Z\"urich, 8092 Z\"urich, Switzerland}
\\\
    \IEEEauthorblockA{\IEEEauthorrefmark{2}Laboratory of Theory and Simulation of Materials, EPFL, 1015 Lausanne, Switzerland}
\\\
}

\IEEEspecialpapernotice{(Invited Paper)}

\maketitle

\begin{abstract}
Full-band atomistic quantum transport simulations based on first
principles are employed to assess the potential of band-to-band
tunneling field-effect-transistors (TFETs) with a 2-D channel material
as future electronic circuit components. We demonstrate that
single-layer transition metal dichalcogenides (TMDs) are not
well-suited for TFET applications. There might, however, exist a great
variety of 2-D semiconductors that have not even been exfoliated yet:
this work pinpoints some of the most promising candidates among them
to realize highly efficient TFETs. Single-layer SnTe, As, TiNBr, and
Bi are all found to ideally deliver ON-currents larger than 100
$\mu$A/$\mu$m at 0.5~V supply voltage and 0.1~nA/$\mu$m OFF-current
value. We show that going from single to multiple layers can boost the
TFET performance as long as the gain from a narrowing band gap exceeds
the loss from the deteriorating gate control. Finally, a 2-D van der
Waals heterojunction TFET is revealed to perform almost as well as the
best single-layer homojunction, paving the way for research in optimal
2-D material combinations.
\end{abstract}


\IEEEpeerreviewmaketitle

\section{Introduction}\label{sec_intro}

\IEEEPARstart{O}{ne} of the main challenges in electronics nowadays
consists in reducing the energy consumption of the myriads of
microprocessors that build the core of consumer products and the
Internet of Things (IoT). In order to make future developments
sustainable, the supply voltage $V_{DD}$ of metal-oxide-semiconductor
field-effect transistors (MOSFETs) - the active components of all
digital logic circuits - should be significantly reduced, while
providing the same current magnitude to maintain high speed
operations. The performance of conventional MOSFETs is however bounded
by the 60~mV/dec theoretical limit of their inverse subthreshold slope (SS) at
room temperature. As a consequence, a gate voltage change $\Delta V_g$
of at least 60 mV must be applied to increase the drive current $I_d$
by one order of magnitude. The ratio between the ON-state current
$I_{ON}$ and the leakage current in the OFF-state, $I_{OFF}$, is
therefore limited. Since the semiconductor industry is rapidly
approaching the maximum $I_{ON}$/$I_{OFF}$ ratio possible in
MOSFETs, novel device types relying on operating mechanisms different
from thermionic emission are under intense investigation \cite{nikonov}. 

Band-to-band tunneling field-effect transistors (TFETs) are promising
candidates for that purpose, as their working principle is based on
the injection of cold instead of hot carriers, effectively creating a
band-pass filter for the high energy electrons that limit the
performance of MOSFETs \cite{ionescu}. The abrupt opening and closing
of a tunneling window with high transmission probability requires an
excellent electrostatic control that is favored by multi-gate,
three-dimensional device structures over planar ones with a single
gate contact. Hence, Silicon and III-V gate-all-around nanowire TFETs
have been demonstrated to break the 60~mV/dec thermal limit of
MOSFETs, but with very low ON-currents \cite{si_tfet,si_iiiv_tfet,siinas}. 
Tunneling transistors with $I_{ON}$'s suitable for practical
applications have been fabricated as well \cite{iiiv_tfet,ingaassb,AsSb}, 
but without simultaneously exhibiting a steep SS. At this point,
delivering a high ON-current and bringing SS well below 60 mV/dec
remains an arduous objective. To the best of our knowledge, a
recently reported InAs/InGaAsSb/GaSb nanowire TFET with
$I_{ON}=10.3$~$\mu$A/$\mu$m and $I_{OFF}=1$~nA/$\mu$m at
$V_{DD}=0.5$~V \cite{best_tfet} represents one of the most impressive
and promising achievements to date. In spite of that, this solution
remains far from reaching the level of performance of Si MOSFETs.

Following the first demonstration of a monolayer MoS$_2$ transistor
\cite{single_layer_mos2}, the research in single- and few-layer
two-dimensional (2-D) semiconductors has undergone a veritable
boom. As part of this on-going effort, TFETs with a 2-D channel
material have started to emerge as an alternative to traditional
semiconductors. The motivation beyond such investigations can be
traced back to the unique features of 2-D materials that can be
leveraged in tunneling devices: (1) the extraordinary gate control
offered by channels with ultra-scaled thicknesses ($<$1 nm), (2) the
variety of the members of the 2-D material family, (3) the large
design space that combinations of 2-D layers opens up, and (4) the
possibility to create heterojunctions with fine-tuned band alignments
and defect-free interfaces thanks to naturally passivated surfaces and
the absence of dangling bonds \cite{van_der_waals}. 

Surfing on this wave, several simulation approaches and design tools
have been developed to shed light on the behavior of 2-D TFETs. In
this context, single-layer transition metal dichalcogenides (TMDs)
\cite{2dtfets,metal_dichalcogenide_tfet1} and black phosphorus
\cite{bp_tfet} TFETs have been extensively studied through computer
models mostly based on the tight-binding (TB) method. Some TMDs,
especially WTe$_2$, were claimed to be able to deliver 
ON-currents larger than 100~$\mu$A/$\mu$m at $V_{DD}=0.5$~V and
$I_{OFF}=1$~nA/$\mu$m. Better device characteristics were predicted by
other theoretical studies for TFETs implementing a van der Waals
hetero-junction as active tunneling region \cite{metal_dichalcogenide_tfet2,drc}. 
Practical devices have been experimentally realized, confirming the
potential of 2-D materials as TFET building blocks
\cite{ge_mos2_tfet,yash_essderc}. It should however be noticed that
the highest ON-current with an inverse subthreshold slope steeper than
60~mV/dec over many orders of magnitudes was achieved by a MoS$_2$-Ge
(2-D to 3-D) heterojunction \cite{ge_mos2_tfet}, leaving open the
question whether 2-D materials alone are appropriate for TFETs. 

To address this issue, we will first revisit the performance of
TFETs based on the most popular single-layer TMDs through advanced
computational modeling. To go one step further, we will then explore
more exotic 2-D materials as possible TFET channels. All of them have
been theoretically proved to be stable in vacuum, their electronic
properties have been analyzed in great details \cite{high_throughput},
but they have not been experimentally studied yet. To complete the
picture, we will extend our investigation to multi-layer devices and
to van der Waals heterojunctions. The focus is set on identifying
components with a high ON-current. 

A quantum transport simulator relying on the Non-Equilibrium Green's
Function (NEGF) formalism and first principles concepts (density
functional theory) will be used to give insight into 
the ``current vs. voltage'' characteristics of all considered
TFETs. This approach avoids the need for empirical parametrizations,
hence enabling the inspection of TFETs made of not-yet exfoliated 2-D
materials. We will show that none of the widely studied TMD monolayers
can reach an $I_{ON}$ larger than approximately 10~$\mu$A/$\mu$m,
while some of the recently discovered 2-D materials significantly
outperform them. We will also demonstrate that going from single- to
few-layer channels can lead to performance enhancement under certain
circumstances. Finally, we will present a 2-D heterojunction TFET that
exhibits a similar ON-current as the best monolayer homojunction. It
is expected that these findings will encourage the search for even
better-suited 2-D materials or material combinations for TFET
channels: spectacular improvements may occur as we move away from
conventional TMDs. 

The paper is organized as follows: in Section \ref{sec_approach}, the
developed \emph{ab initio} simulation approach is introduced. Simulation
results of 2-D TFETs are given in Section \ref{sec_results},
starting with standard TMDs, then novel 2-D materials, multi-layer
channels, and van der Waals heterojunctions to finish. Conclusions are
drawn in Section \ref{sec_conclusion}.

\section{Simulation approach}\label{sec_approach}

The simulation workflow used in this paper starts with the
calculation of the single-particle electronic wavefunctions in the
primitive unit cell of the investigated 2-D materials. This is 
accomplished with the help of density functional theory (DFT)
\cite{kohn} as implemented in the VASP package \cite{vasp}. The
generalized gradient approximation (GGA) of Perdew, Burke, and
Ernzerhof (PBE) \cite{PBE} is utilized in most cases since it has been 
demonstrated to give a band gap value very close to the experimental
one for single-layer MoS$_2$ \cite{vasp_gga}. The only exception is
black phosphorus (1-, 2-, and 3-layer), where the HSE06 hybrid
functional \cite{HSE06} is employed due to its better
reproduction of the experimental band gap as compared to PBE that
severely underestimates it. It should be pointed out
that there is no consensus on the band gap of most 2-D
materials. Supposedly more accurate G$_0$W$_0$ calculations predict,
for example, $E_g$=2.78 eV for MoS$_2$ \cite{louie}. Such large
corrections could have a negative influence on the results presented in
this paper that can be seen as best case scenarios.

An 500~eV plane-wave cutoff energy, a $25\times25\times1$
Monkhorst-Pack $k$-point grid, and 20 $\AA$ out-of-plane vacuum
separation are applied to every DFT simulation. Spin-orbit
coupling is neglected. The convergence criteria is set to less than
$10^{-3}$ eV/$\AA$ forces acting on each ion and a total energy
difference smaller than $10^{-3}$~eV between two consecutive
self-consistent field iterations. In the case of multilayer
structures, van der Waals interactions are included through the DFT-D2
method of Grimme \cite{vdw}. Heterojunctions are modeled in a common
unit cell with the average of the lattice parameters of the
individual layers. The resulting stress (compressive for one layer,
tensile for the other) does not exceed 2\%.

\begin{figure}[!t]
\centering
\includegraphics[width=\linewidth]{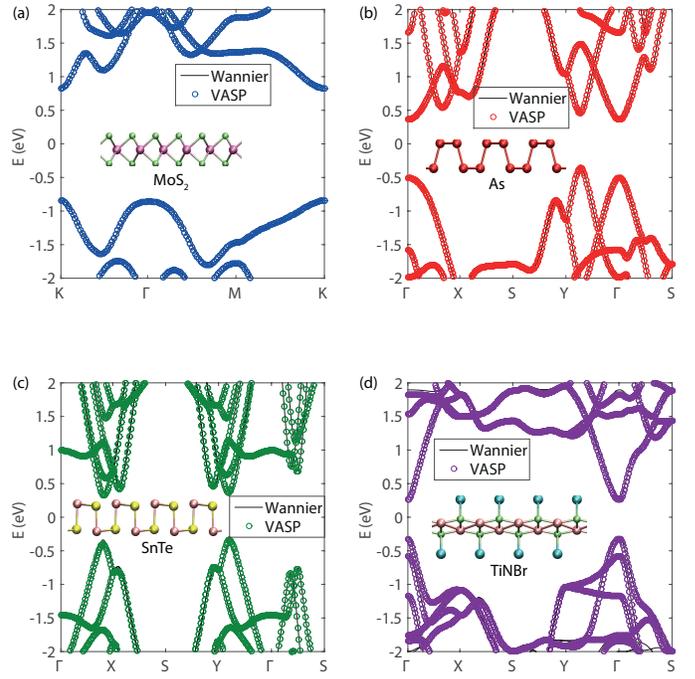}
\caption{Bandstructures of selected single-layer materials obtained
  from VASP (circles) and from sparse MLWF Hamiltonians after
  applying a cutoff radius of about 15 $\AA$ to the interactions between
  different Wannier functions (lines): (a) MoS$_2$, (b) As, (c) SnTe,
  (d) TiNBr. The MLWF Hamiltonians capture all relevant features for
  transport calculations with high fidelity. The corresponding crystal
  structures are shown as insets.}
\label{fig:bandstructures}
\end{figure}

The Bloch wavefunctions produced by VASP for a structure without any
applied electric field are then transformed into a
set of maximally localized Wannier functions (MLWFs) by the Wannier90
tool \cite{wannier}. The Hamiltonian matrix elements in this basis
correspond to interactions between well-localized wavefunctions
typically centered on ions. Due to this localization, interactions
between Wannier functions separated by a distance larger than a
pre-defined cutoff radius can be safely neglected. This cutoff radius
is determined on a case-by-case basis to minimize the computational
burden, while maintaining a high accuracy. Its typical value lies in
the range of 15 \AA. The Hamiltonian matrices of 80~nm long TFETs are
built up from the retained matrix elements. They exhibit a
block-tri-diagonal structure with a tight-binding-like sparsity
pattern. They can be loaded into a ballistic quantum transport
solver relying on the Non-Equilibrium Green's Function formalism
\cite{omen}. This ``up-scaling'' procedure is further detailed in
Ref.~\cite{mos2_prb} and Appendix~A of Ref.~\cite{mythesis} for
homogeneous materials and in Ref.~\cite{edl} and Appendix~B of
Ref.~\cite{mythesis} for heterojunctions. The accuracy of the method
is confirmed in Fig.~\ref{fig:bandstructures} for four different
single-layer 2-D materials,  MoS$_2$, As, SnTe, and TiNBr. Still,
should the bandstructure of these crystals be modified by a strong
electric field, it is not clear whether the MLWF Hamiltonian would
automatically capture these changes. 

The NEGF quantum transport simulator solves the Schr\"odinger- and
Poisson equations self-consistently in the presence of electrostatic
boundary conditions imposed by the metallic gates. We construct
80~nm long simulation domains with $L_g=40$~nm top and bottom
gate contacts separated from the channel materials by 3~nm thick
high-$\kappa$ HfO$_2$ layers with $\varepsilon_R=20$ and an equivalent
oxide thickness EOT$=0.6$ nm. The source and drain extensions measure
approximately $L_s=13$~nm and $L_d=26$~nm, the rounding coming from
the fact that the total TFET length has to be an integer multiple of the 
primitive unit cell width $\Delta_{uc}$. These regions are doped with a
$N_A=5\times10^{13}$~cm$^{-2}$ and $N_D=5\times10^{12}$~cm$^{-2}$
acceptor and donor concentrations, respectively, that are modeled as 
uniformly screened background charges in the Poisson equation. Such an
ideal configuration could only be imagined in the context of
electrostatic doping resulting from additional metal contacts or from
the transfer of charges from another 2-D layer stacked with the actual
one \cite{lu}. As compared to a more realistic atomistic model
explicitly accounting for chemical doping or substitutional
impurities, the present scheme offers a ``best-case'' scenario that is
suitable to study the performance limit of a given design. Going
beyond this approximation is out of the scope of this paper.

\begin{figure}[!t]
\centering
\includegraphics[width=\linewidth]{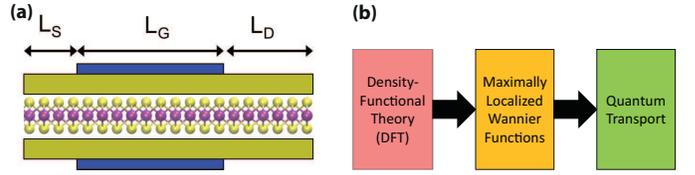}
\caption{(a) Schematic view of the considered double-gate TFETs with
  monolayer 2-D materials as channels. (b) Flow chart of the
  simulation approach based on density functional theory, maximally
  localized Wannier functions, and NEGF quantum transport.}
\label{fig:device_and_workflow}
\end{figure}

Only $n$-type TFETs are simulated here. Their different source and
drain lengths and doping concentrations ensure sharp changes in the
electrostatics at the source-channel interface, which maximizes the
tunneling probability there, and a more smoothly varying potential on
the drain side, thus avoiding tunneling leakage currents. The 2-D
materials are treated as flakes with one periodic direction, the
out-of-plane one, that induces a $k$-dependence of all computed
physical quantities. A total of 15 $k$-points between 0 and $\pi$ have
been found sufficient to represent this direction. The dielectric
constants of the TMDs are taken from Ref.~\cite{eps}, whereas those of
novel 2-D materials have been computed according to Ref.~\cite{sohier}. 
The source-to-drain voltage of the TFETs, $V_{ds}$, is set to 0.5~V and
the gate voltage $V_{gs}$ is swept from 0 to 0.5~V. The workfunction
of the gate metals is adjusted in each case to set the OFF-current at
$V_{gs}$=0 V to $I_{OFF}$=0.1 nA/$\mu$m. Figure \ref{fig:device_and_workflow} 
depicts the schematics of the studied tunneling devices as well as a
summary of the simulation workflow. 

\section{Results and discussion}\label{sec_results}

To illustrate the potential of 2-D materials as active components of
future TFETS, the investigated crystals have been divided into four
generic categories: 
\begin{itemize}
 \item single-layer transition metal dichalcogenides: MoS$_2$,
   MoSe$_2$, MoTe$_2$, WS$_2$, WSe$_2$, WTe$_2$ (all of the trigonal
   prismatic form), and octahedral HfSe$_2$; 
 \item novel single-layer 2-D materials with small direct band gaps,
   as predicted by the theoretical study from
   Ref.~\cite{high_throughput}: AlLiTe$_2$, As, Bi, CrSe$_2$, SnTe,
   TiNBr, and TiNCl; 
 \item single-, double-, and triple-layer black phosphorus; 
 \item a MoTe$_2$-SnS$_2$ van der Waals heterojunction composed of two
   single-layer crystals.
\end{itemize}
These four groups will be presented and analyzed one by one in the
following subsections.

\subsection{Transition metal dichalcogenides}

\begin{figure}[!t]
\centering
\includegraphics[width=\linewidth]{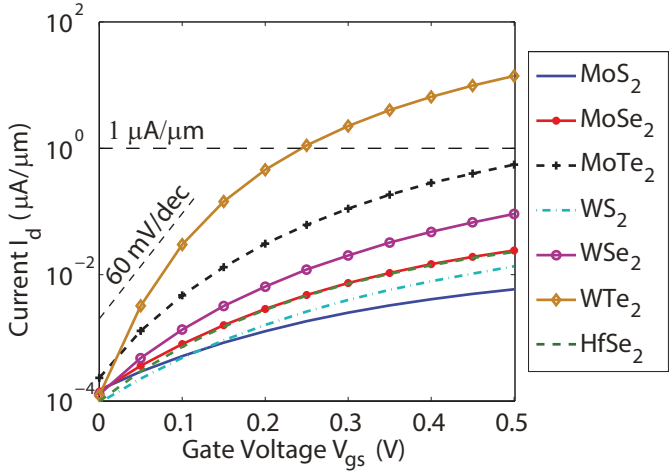}
\caption{Transfer characteristics $I_d-V_{gs}$ at $V_{ds}=0.5$ V of
  conventional single-layer 2-D materials: MoS$_2$ (solid line),
  MoSe$_2$ (solid line with dots), MoTe$_2$ (dashed line with
  crosses), WS$_2$ (dashed-dotted line), WSe$_2$ (solid line with
  circles), WTe$_2$ (solid line with diamonds), and HfSe$_2$ (dashed
  line). The 60 mV/dec slope and the 1 $\mu$A/$\mu$m mark are
  indicated.}
\label{fig:TMD_results}
\end{figure}

Molybdenum- and tungsten-based single-layer (SL) transition metal
dichalcogenides (TMDs) are all direct gap semiconductors with WSe$_2$
exhibiting the largest and WTe$_2$ the smallest theoretical band gap
of 1.8 and 0.92~eV, respectively. Note that the trigonal prismatic
form of WTe$_2$, the semiconducting allotrope investigated in this
work, is metastable. Reaching high tunneling probabilities requires a 
small band gap to reduce the tunneling length and small electron/hole
effective masses to minimize the decay rate $\kappa$ along the tunneling
path. As Fig.~\ref{fig:TMD_results} shows, none of the conventional SL 
2-D materials fully satisfies these essential criteria. The simulated
ON-currents are smaller than 1 $\mu$A/$\mu$m in all cases, except for
WTe$_2$, where $I_{ON}$ reaches 14 $\mu$A/$\mu$m. This result is
consistent with WTe$_2$ having the smallest direct band gap and the
smallest average electron/hole effective mass in this group, but not
sufficient to challenge Si MOSFETs. It should also be emphasized that
our drive currents $I_d$ are much smaller than those of
Ref.~\cite{2dtfets}. This could be due to the simulation approach,
empirical tight-binding (TB) in Ref.~\cite{2dtfets} vs. DFT
here. There is no guarantee that with a TB method, the imaginary band
dispersion responsible for intra- and inter-band tunneling is
accurately reproduced, while it is an integral part of DFT
bandstructures and their MLWF representation. 

Among the considered TMDs, HfSe$_2$ is the one with the smallest band
gap, 0.76 eV, but it is indirect, leading to a reduced transmission
probability. Although we only perform ballistic simulations, indirect
tunneling can still occur if the transport direction $x$ is aligned
with the wavevector that connects the minimum of the conduction band
(CBmin) and the maximum of the valence band (VBmax) of the 2-D
material under consideration. This is not the case here since CBmin
(VBmax) of HfSe$_2$ is located at the $M$ ($\Gamma$) point and the
transport direction $x$ follows the $\Gamma$-$K$ axis of the Brillouin
Zone. Aligning $x$ with the $\Gamma$-$M$ line would require larger
unit cells and increased computational burden. For this reason, 
this case has not been examined here. 

\begin{figure}[!t]
\centering
\includegraphics[width=0.8\linewidth]{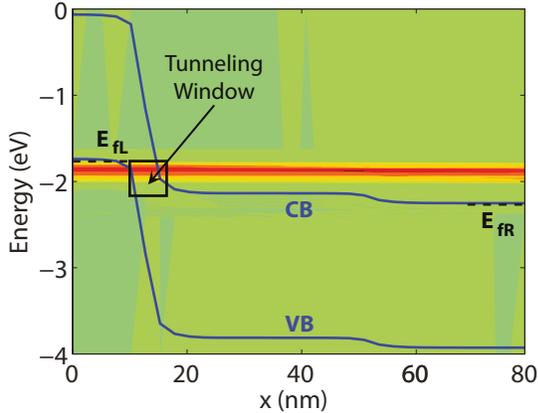}
\caption{Spectral current in the OFF-state of the MoS$_2$
  TFET ($V_{gs}$=0 V, $V_{ds}$=0.5 V). Red indicates high current
  concentrations, green no current. The blue lines refer to the band
  edges, $E_{fL}$ and $E_{fR}$ to the source and drain Fermi levels,
  respectively. The black square represents the tunneling window where 
  electrons in the valence band of the source can be ballistically
  transferred to the conduction band of the channel.}
\label{fig:MoS2_spectral_and_geoms}
\end{figure}

Apart from the small ON-currents, it can be observed that almost all
TMDs have inverse subthreshold slopes larger than 60~mV/dec despite their
excellent electrostatic properties. The only exception is WTe$_2$,
whose SS surpasses the thermal limit. The reason for this behavior is
clarified in Fig.~\ref{fig:MoS2_spectral_and_geoms} that shows the
spectral current in the OFF-state of the MoS$_2$ TFET. To achieve
$I_{OFF}$=0.1 nA/$\mu$m, a tunneling window with a significant width
must already be available: due to the low tunneling probabilities of
MoS$_2$ and most other TMDs, even the minuscule imposed OFF-current
requires a gate voltage much higher than the one at which the valence
band of the source is perfectly aligned with the conduction band of
the channel. A sub-thermionic SS can only be expected when this
condition is satisfied, that is, in the present case, at negative
voltages and not relevant current levels. 

The band gaps, electron and hole effective masses, and the computed
ON-current values are summarized in Table \ref{tab:all_data} situated
at the end of the manuscript. 

\subsection{Novel single-layer 2-D semiconductors}

From the newly generated database of all potentially existing 2-D
materials (more than 1,800 compounds) \cite{high_throughput}, we
singled out a few examples with very promising parameters for
application as TFET channels. The components with small and direct
band gaps have been selected. Still, $E_g$ should not be smaller than 
0.5~eV, otherwise turning off the TFET would not be possible at
$V_{ds}=0.5$~V. The material from this group with the largest band gap
is AlLiTe$_2$ with 0.91~eV, which is comparable to WTe$_2$, while the
smallest $E_g$ is obtained for Bi with 0.55~eV. Besides them, the
following single-layer crystals are considered: As, CrSe$_2$, SnTe,
TiNBr, and TiNCl.

\begin{figure}[!t]
\centering
\includegraphics[width=\linewidth]{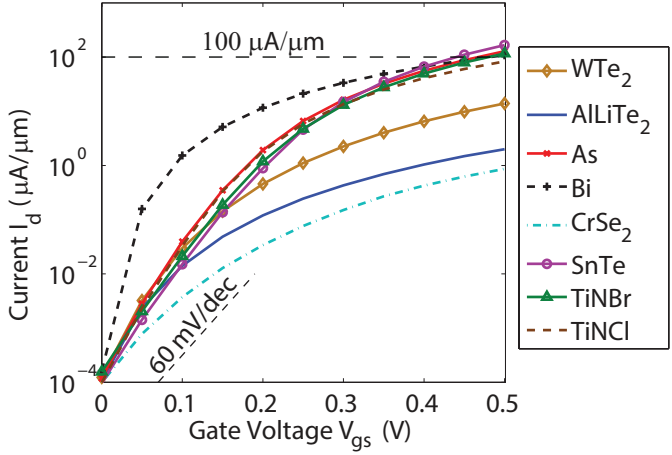}
\caption{Transfer characteristics $I_d-V_{gs}$ at $V_{ds}=0.5$~V of
  the novel single-layer 2-D materials considered in this work:
  AlLiTe$_2$ (solid line), As (solid line with crosses), Bi (dashed
  line with crosses), CrSe$_2$ (dashed-dotted line), SnTe (solid line
  with circles), TiNBr (solid line with triangles), and TiNCl (dashed
  line). They are compared to WTe$_2$ (solid line with diamonds).}
\label{fig:novel}
\end{figure}

The transfer characteristics $I_d-V_{gs}$ at $V_{ds}=0.5$~V of the
TFETs made of these materials are shown in Fig.~\ref{fig:novel}. This
time, most of the simulated devices reach ON-currents in the range of
100~$\mu$A/$\mu$m or more, with the exception of the material with the
largest gap, AlLiTe$_2$, and the largest effective masses,
CrSe$_2$. The highest ON-current is found for SnTe, 166 $\mu$A/$\mu$m,
one order of magnitude above WTe$_2$, the best performing TMD. The
devices with more than 100~$\mu$A/$\mu$m ON-currents display clear
sub-thermionic operation in the subthreshold regime, with slopes
steeper than 60~mV/dec in each case. Bi has the lowest SS thanks to
its very low band gap and effective masses. To be able to switch off
all these TFETs, the source and drain doping concentrations of the Bi,
TiNCl, and TiBrN structures had to be decreased by a factor of 5, 2, and
1.25, respectively. This explains why the current increase in the Bi
TFET rapidly saturates compared to the other 2-D materials. 

\begin{figure}[!t]
\centering
\includegraphics[width=\linewidth]{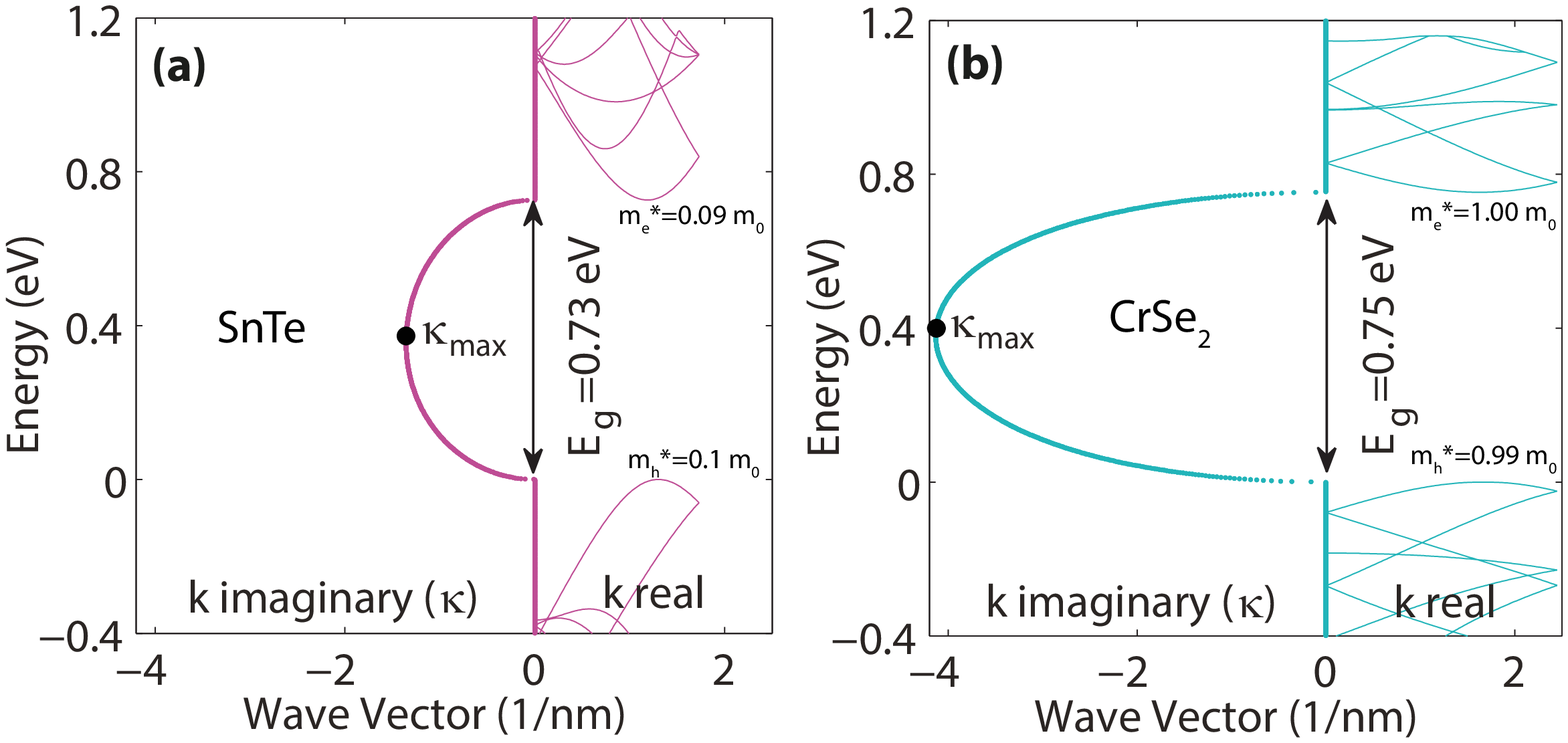}
\caption{Complex bandstructure of SL (a) SnTe and (b) CrSe$_2$. The
  left axis refers to the imaginary part of the band (energy-resolved
  decay rate $\kappa$) multiplied by a minus sign, the right axis to
  its real component $k$. The maximum value of $\kappa$,
  $\kappa_{max}$, is indicated for both materials.}
\label{fig:kappa}
\end{figure}

The influence of the effective masses on the device performance is
illustrated in Fig.~\ref{fig:kappa}, where the complex bandstructure
of SnTe and CrSe$_2$ is displayed. The band gaps of these 2-D materials
are almost identical, 0.73 eV for SnTe and 0.75 eV for CrSe$_2$. The
effective masses, however, differ by a factor of 10: both the electron
and hole effective masses of SnTe are approximately equal to 0.1
m$_0$, while they amount to about 1.0 m$_0$ in CrSe$_2$. Hence, the
maximum decay rate $\kappa_{max}$ of the electronic wavefunction
inside the band gap of these semiconductors becomes quite different,
1.36 1/nm for SnTe vs. 4.1 1/nm for CrSe$_2$, roughly a factor
$\sqrt{10}$ difference, which corresponds to the square root of the
CrSe$_2$/SnTe effective mass ratio. A high $\kappa_{max}$ means that
the wavefunction of an electron injected into the valence band of the
source extension of a TFET rapidly diminishes as it moves towards the
conduction band of the channel, thus reducing the tunneling
probability, regardless of the band gap value \cite{tunneling_exponential}. 
This is the main reason why CrSe$_2$ is outperformed by SnTe as well
as by every other 2-D material considered in this sub-section. 

\subsection{Single- and few-layer black phosphorus}

Exploring novel single-layer 2-D semiconductors is one promising
possibility to improve the TFET performance, another one could be the
introduction of additional layers to the channel. We will therefore
investigate here whether going from single- to few-layer structures
offers any benefit. Black phosphorus (BP) is chosen for this task, as
its layer-dependent band gap, high mobility, and high anisotropy of
the electron and hole effective masses are already well-known
\cite{BP_anisotropy}. The mass anisotropy can be exploited by
orienting the transport axis with the small mass direction, hence
combining the advantage of a high electron velocity together with a
high density-of-states \cite{szabo_bp}. A previous theoretical study
based on the tight-binding method and an assumed band gap of
$E_{g,SL}$=1.4 eV for the single-layer compound reported ON-state
currents larger than 1 mA/$\mu$m at $V_{DD}$=0.5 V for multi-layer
TFETs \cite{bp_tfet}. Here, $E_{g,SL}$=1.61 eV and DFT are used. 

\begin{figure}[t]
\centering
\includegraphics[width=\linewidth]{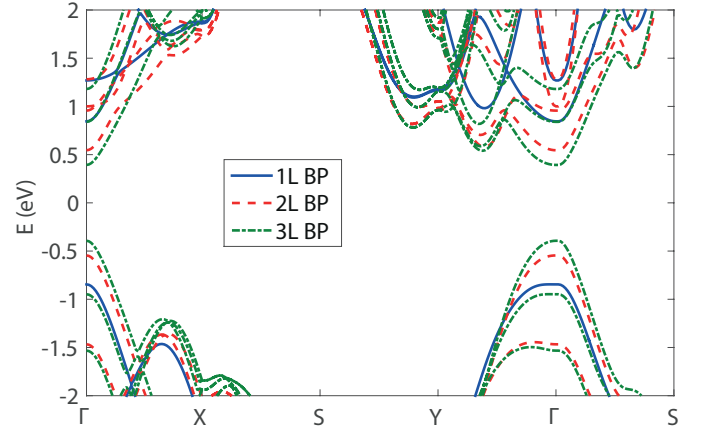}
\caption{Bandstructure of single- (solid lines), double- (dashed lines),
  and tripe-layer (dashed-dotted lines) black phosphorus as computed
  with DFT and HSE06 functional \cite{HSE06}.}
\label{fig:BP_bs}
\end{figure}

With this band gap value and relatively large effective masses
($m_e$=0.17 for electrons and $m_h$=0.16 for holes), SL BP cannot
offer high tunneling currents. Increasing the channel thickness to 2
layers lowers the gap to 1.09 eV, while for 3 layers, it further
decreases down to 0.79 eV. The bandstructures of BP with 1 to 3 layers
are shown in Fig.~\ref{fig:BP_bs}. In each configuration, the band gap
remains direct and is located at the $\Gamma$ point, contrary to most
TMDs, where a direct-indirect transition takes place when going from a
single- to a multi-layer crystal. Transport occurs
along the $X$ direction: only the light electrons and holes of the
$\Gamma-X$ path (all in the range between 0.14 and 0.18 m$_0$)
contribute to the tunneling process in the simulated devices. 

\begin{figure}[t]
\centering
\includegraphics[width=\linewidth]{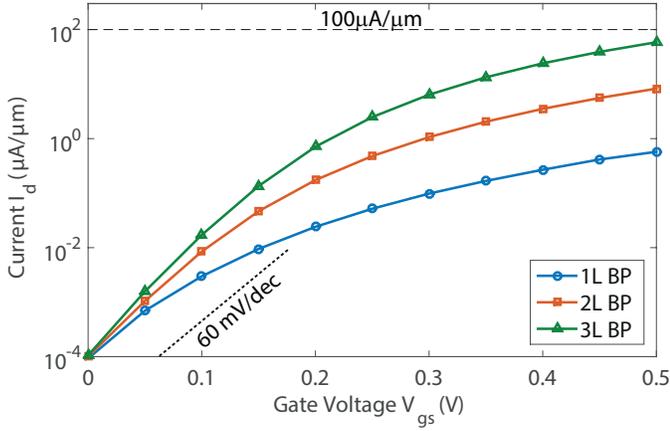}
\caption{Transfer characteristics $I_d-V_{gs}$ at $V_{ds}=0.5$~V of
  single- (solid line with circles) double- (solid line with square),
  and triple-layer (solid line with triangles) BP TFETs.}
\label{fig:BP_iv}
\end{figure}

Thicker channels come at the expanse of a reduced electrostatic
control. However, as indicated by the transfer characteristics of
single-, double-, and triple-layer BP TFETs in Fig.~\ref{fig:BP_iv},
the gain from the smaller gaps overcomes the loss caused by the
deterioration of the electrostatics. While the ON-current of SL BP
only reaches 0.57~$\mu$A/$\mu$m, it increases to 8.3~$\mu$A/$\mu$m in a
double-layer and to 58.6~$\mu$A/$\mu$m in a triple-layer BP TFET. This
enhancement can be mainly attributed to the narrowing of the band gap.
The effective mass variations are smaller than 10~\%, whereas the 
band gap decreases by a factor of two between the single- and
triple-layer crystal. These results are better than those obtained for
conventional TMDs, but not for SnTe and other novel 2-D semiconductors.

It is not clear whether this ON-current improvement will continue for
even thicker channels, and if so, what is the optimal thickness. At
one point, by adding more and more layers, the band gap will approach
0.5 eV and the OFF-state leakage will start to exponentially
increase. At the same time, the gate control will gradually lose its
efficiency, which will affect the subthreshold region. Simulating
thicker channels is out of the scope of this work as the computational
burden grows very quickly with the cross-section of the device. 

\begin{figure}[!t]
\centering
\includegraphics[width=\linewidth]{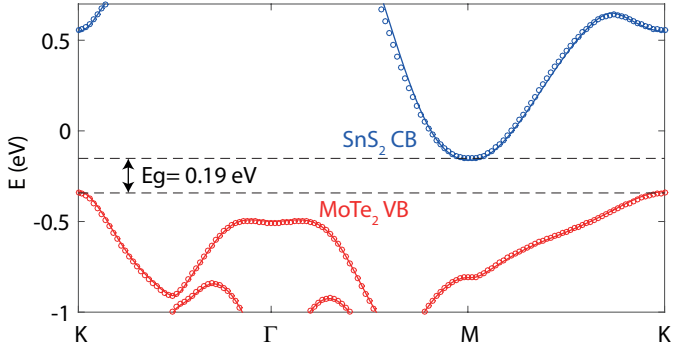}
\caption{Band structure of the MoTe$_2$-SnS$_2$ van der Waals
  heterojunction. The highest valence bands are formed by the MoTe$_2$
  layer, the lowest conduction bands belong to the SnS$_2$
  layer. Circles mark the results calculated with VASP, lines are from
  the MLWF Hamiltonian.}
\label{fig:mote2_sns2_bandstructure}
\end{figure}

\subsection{MoTe$_2$-SnS$_2$ van der Waals heterojunction}

\begin{figure}[t]
\centering
\includegraphics[width=\linewidth]{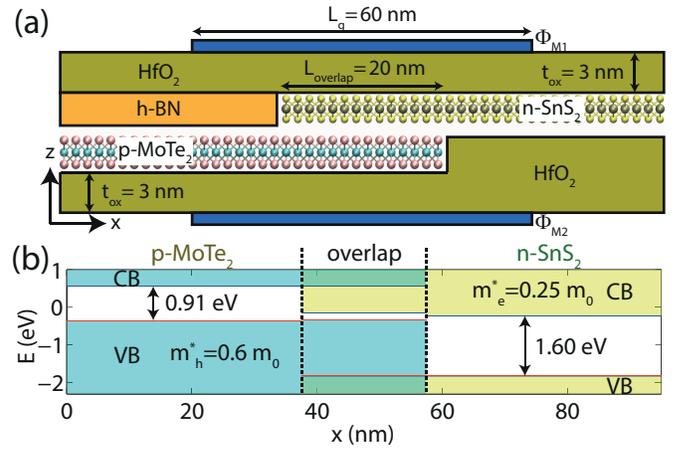}
\caption{(a) Schematic view of the considered double-gate
  MoTe$_2$-SnS$_2$ hetero-TFET. (b) Conduction and valence band
  profiles of the device along the transport axis under flat band 
  conditions.}
\label{fig:mote2_sns2_device}
\end{figure}

As last TFET concept, a MoTe$_2$-SnS$_2$ van der Waals heterojunction
is proposed. The bandstructure of this material combination is
depicted in Fig.~\ref{fig:mote2_sns2_bandstructure}, its device
arrangement schematized in Fig.~\ref{fig:mote2_sns2_device}~(a). A
similar TFET was previously discussed in Ref.~\cite{edl}. Its
structure is slightly different from the one in
Fig.~\ref{fig:device_and_workflow}. The source extension is formed by
a $p$-doped MoTe$_2$ monolayer with a $N_A$=$10^{13}$ cm$^{-2}$
acceptor concentration, the drain by a $n$-doped SL SnS$_2$
with a $N_D=10^{13}$~cm$^{-2}$ donor concentration. The two layers
overlap in the middle of the device over a 20 nm long region. The top
and bottom gate contacts extend on both sides by 20 nm. The motivation
behind the elongated gates is the desire to avoid tunneling paths at
the edges of the overlap area and to orientate the transfer of electrons
from one layer to the other along the vertical $z$ direction. One layer
of hexagonal Boron Nitride (BN) is deposited on top of the
free-standing part of MoTe$_2$ to eliminate strong potential
variations close to the interface edges. In effect, BN has a
dielectric permittivity closer to both MoTe$_2$ and SnS$_2$ than
HfO$_2$, the surrounding oxide. We use the same supply voltage
$V_{DD}=0.5$ V as before and set the OFF-current to $I_{OFF}$=0.1
nA/$\mu$m by adjusting the gate workfunctions. This makes the
established results comparable to TMDs, novel 2-D materials, and 
few-layer BP.

\begin{figure}[t]
\centering
\includegraphics[width=\linewidth]{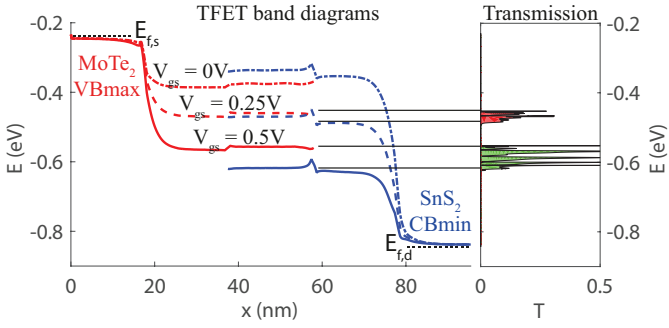}
\caption{Band diagram and energy-resolved transmission probability in
  the MoTe$_2$-SnS$_2$ TFET at various gate voltages. The valence band
  maximum (VBmax) of MoTe$_2$ and the conduction band minimum (CBmin)
  of SnS$_2$ are represented. The horizontal black lines mark the
  edges of the tunneling windows at 0.25 and 0.5 V gate voltages,
  encompassing the red and green transmission profiles, respectively.}
\label{fig:mote2_sns2_bands}
\end{figure}

MoTe$_2$ and SnS$_2$ are two semiconductors with relatively large band
gaps for tunneling applications, 0.91 and 1.6 eV, respectively, but
their stacking forms a staggered band alignment with a tunneling gap
of only 0.19 eV. The band diagram of the MoTe$_2$-SnS$_2$ van der
Waals heterojunction under flat band conditions is shown in
Fig.~\ref{fig:mote2_sns2_device}~(b). The TFET operates by grounding
the bottom gate and sweeping the top one. A vertical electric field is
created by this process. In the ON-state, it pushes down the
conduction band edge of SnS$_2$ below the valence band maximum of
MoTe$_2$ because the potential close to the top layer is better
controlled by the gate contact. This mechanism can be visualized in
Fig.~\ref{fig:mote2_sns2_bands}. At $V_{gs}=0$~V the valence band
maximum (VBmax) of MoTe$_2$ is at a lower energy than the conduction
band minimum (CBmin) of SnS$_2$ in the active regions: there is no
tunneling path. As the gate voltage increases, the CBmin of SnS$_2$
moves downwards faster than the VBmax of MoTe$_2$, closing the 0.19 eV
gap between both materials. At $V_{gs}$=0.25 V, a narrow tunneling
window appears in the middle of the device, but most of the
current flows at the right edge of the overlap region, from
MoTe$_2$ into the free-standing SnS$_2$ layer (point tunneling). In
the ON-state, a large tunneling window is formed in the overlap area
(line tunneling). It leads to a high ON-current $I_{ON}$=130
$\mu$A/$\mu$m, the tunneling length being the distance between both
semiconducting layers, i.e. less than 0.5 nm. The sharp peaks observed
in the ballistic transmission functions plotted in
Fig.~\ref{fig:mote2_sns2_bands} result from the presence of resonance
and anti-resonance states. The latter either enhance or deteriorate
the transmission probability. The inclusion of a dissipative
scattering mechanism such as electron-phonon interactions would either
eliminate them or broaden their spectral width.  

\begin{figure}[t]
\centering
\includegraphics[width=\linewidth]{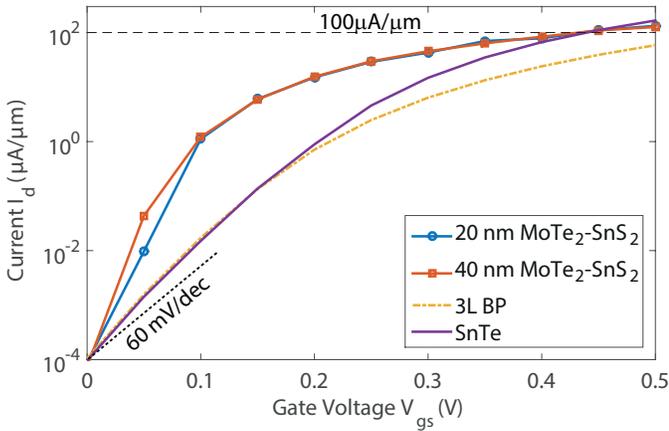}
\caption{Transfer characteristics $I_d-V_{gs}$ at $V_{ds}$=0.5 V of
  the MoTe$_2$-SnS$_2$ heterojunction TFETs with 20 (solid line with
  circles) and 40 nm (solid line with squares) overlap
  lengths. Triple-layer black phosphorus (dashed line) and SnTe (solid
  line) are shown for comparison.}
\label{fig:mote2_sns2_iv}
\end{figure}

The role of the overlap length has also been investigated by
simulating a TFET where this parameter is increased to 40 nm, 
twice as large as before. The original and modified devices deliver
almost identical currents, as shown in Fig.~\ref{fig:mote2_sns2_iv}. 
This suggests that the tunneling current is not homogeneously
distributed over the overlap region. Indeed, a large contribution
comes from the current at the right edge of the interface. Still, the
constructed heterojunction achieves an ON-current $I_{ON}$=130
$\mu$A/$\mu$m, which is higher than the TFET relying on
triple-layer BP and comparable to what is found for the studied novel
2-D semiconductors, as summarized in Table \ref{tab:all_data}. These
results are very encouraging since the playground for van der Waals
heterojunctions is almost infinite (material type and number of
layers), especially if it is recalled that 1,800 2-D crystals might
exist \cite{high_throughput}. 

\section{Conclusion}\label{sec_conclusion}
\label{sec:conclusion}

By performing atomistic quantum transport simulations based on first
principles, we have demonstrated that single-layer transition metal
dichalcogenides are not competitive as TFET channel materials. None of
them, except for WTe$_2$, has a ON-current that exceeds 1
$\mu$A/$\mu$m at a supply voltage $V_{DD}$=0.5 V and $I_{OFF}$=0.1
nA/$\mu$m. However, several novel 2-D single-layer semiconductors that
have just recently started to attract some attention could play a
prominent role in future TFETs. Some of them such as SnTe, As, TiNBr,
and Bi are predicted to deliver $I_{ON}>$100~$\mu$A/$\mu$m. Extending
the range of search to devices made of few-layer crystals may further
improve the situation if the considered semiconductors have a low and
direct band gap, small electron and hole effective masses, and their
thickness remains narrow enough to maintain an excellent
electrostatic control via the gate contact. A bright future can be
envisioned for TFETs with a 2-D channel material as the design space
is huge and components with significantly better performance than
conventional TMDs have already been identified. 

Nevertheless, there is a practical lower limit on the band gap in
homojunction TFETs. It is imposed by the supply voltage. The materials 
presented here, in particular Bi, are already very close to this
limit. Finding a 2-D semiconductor with a similar direct band
gap, but lower electron and hole effective masses might lead to
improved performance, but might not be sufficient to surpass what van
der Waals heterojunctions can provide. A hetero-TFET with vertical
transport can still be turned off with arbitrarily small staggered
gap values, as long as the band gaps of the individual layers are
reasonably large. The simulated MoTe$_2$-SnS$_2$ device is made
of well-known 2-D compounds. In spite of that, it performs almost as
well as the best homojunction TFETs using novel 2-D semiconductors as
their channel. A better engineered hetero-TFET combining layers with
smaller effective masses and a lower tunneling gap is likely to
outperform all the examples explored in this paper. The next step is
therefore to determine the most suitable material combinations for
heterojunction TFETs from the large database of 2-D semiconductors. 

\begin{table}
\centering
\begin{tabular}{|c|c|c|c|c|}
\hline
& $E_g$~(eV)& $m^{*}_e$~($m_0$) & $m^{*}_h$~($m_0$) & $I_{\textnormal{ON}}$~($\mu$A/$\mu$m) \\ \hline
MoS$_2$ & 1.67 & 0.46 & 0.56 & $5.9\times 10^{-3}$ \\ \hline
MoSe$_2$ & 1.43 & 0.53 & 0.63 & $2.4\times10^{-2}$ \\ \hline
MoTe$_2$ & 1.06 & 0.54 & 0.68 &  $5.5\times10^{-1}$\\ \hline
WS$_2$ & 1.80 & 0.30 & 0.38 & $1.4\times10^{-2}$ \\ \hline
WSe$_2$ & 1.53 & 0.33 & 0.43 & $9.1\times10^{-2}$ \\ \hline
WTe$_2$ & 0.92 & 0.28 & 0.39 & 13.95 \\ \hline
*HfSe$_2$ & 0.76 & 0.19 & 0.14 & $2.3\times10^{-2}$ \\ \hline
AlLiTe$_2$& 0.91 & 0.15 & 0.36 & 1.99 \\ \hline
As & 0.81  & 0.08 & 0.08 & 127.9 \\ \hline
Bi & 0.55 & 0.04 & 0.03 & 108.0 \\ \hline
CrSe$_2$ & 0.75 & 1.00 & 0.99 & $8.6\times10^{-1}$ \\ \hline
SnTe & 0.73 & 0.09 & 0.10 & 166.3 \\ \hline
TiNBr & 0.63 & 0.18 & 0.17 & 116.7 \\ \hline
TiNCl & 0.58 & 0.18 & 0.17 & 83.2 \\ \hline
1L-BP & 1.61 & 0.17 & 0.16 & $5.7\times10^{-1}$ \\ \hline
2L-BP & 1.09 & 0.18 & 0.16 & 8.3 \\ \hline
3L-BP & 0.79 & 0.16 & 0.14 & 58.6 \\ \hline
MoTe$_2$-SnS$_2$ & 0.19  & 0.25 & 0.6 & 130 \\ \hline
\end{tabular}
\begin{flushleft}
* Indirect band gap
\end{flushleft}
\caption{Summary of the band gaps ($E_g$), electron ($m^{*}_e$)
  and hole effective masses ($m^{*}_h$) of the 2-D semiconductors 
  considered in this work, and computed TFET ON-currents
  ($I_{ON}$).}
\label{tab:all_data}
\end{table}

\section{Acknowledgments}\label{sec:ack}

This work was supported by the MARVEL NCCR of the Swiss National
Science Foundation and by ETH Grant ETH-32 15-1. We acknowledge PRACE
for awarding us access to Piz Daint at CSCS under Project pr28. 


%





\ifCLASSOPTIONcaptionsoff
  \newpage
\fi



%

\bibliographystyle{IEEEtran}
\bibliography{IEEEabrv,references_long}


%








\end{document}